\documentclass{moriond}

\def\be{\begin{equation}}
\def\ee{\end{equation}}
\def\bea{\begin{eqnarray}}
\def\eea{\end{eqnarray}}

\newcommand{\Photo}{}

\begin{document}
\vspace*{4cm}
\title{Interpretation of the binary black hole mass spectrum}

\author{Ilya Mandel}
\address{School of Physics and Astronomy, Monash University, Clayton, Victoria 3800, Australia}
\address{OzGrav, Australian Research Council Centre of Excellence for Gravitational Wave Discovery, Australia}

\maketitle\abstracts{
This is a summary of an invited talk given at the Moriond Gravitation meeting on March 31, 2025.  I touch on some of the practical challenges of measuring the mass spectrum of merging binary black holes through their gravitational-wave signatures.  I then describe my take on the current state of interpreting the observed binary black hole mass spectrum from the perspective of models for the formation of these sources. I conclude that meaningful progress must rely on the combination of gravitational-wave observations and a broad range of electromagnetic observations of massive stellar binaries at earlier stages of their evolution.  This is my very personal and necessarily brief take on the current state of the field and does not aspire to the balance or completeness of a review.
}

\section{How well do we observe the binary black hole mass spectrum?}

Let us begin by discussing how we actually measure the black hole population properties, including masses, and how much faith we can place in these measurements.  

Firstly, there is a question of detecting binary black holes.  A great deal of work by members of the LIGO-Virgo-Kagra collaboration over the years -- and, more recently, several competing teams -- has produced a variety of search pipelines.  Broadly, these pipelines share the same goals: to detect gravitational-wave signatures in noisy detector data.  However, they differ in the details of just how they attempt to distinguish signals from noise.  These differences stem from decisions regarding which signals to focus on (e.g., lighter or heavier black holes? similar companion masses or more extreme mass ratios, too?  black holes with aligned or precessing spins?), how to handle non-stationary, non-Gaussian noise artefacts, and precisely what detection statistics to use for ranking candidates.  The consequence is that while the confidently detected candidates typically significantly overlap between search pipelines, there can be some disagreement about the ranking of more marginal candidates, leading to mismatches between the released event catalogs\cite{GWTC3,Nitz:2021-4OGC,Wadekar:2023}.  While this could lead to differences in the inferred mass spectrum, such differences are likely to be of limited impact.

Secondly, there is the question of parameter estimation on detected signals.  Again, the broad approach has been developed by the community over approximately two decades and is now very robust.  It typically involves Bayesian inference on the parameters of the signal given a gravitational waveform model and an additive noise model.  Waveform models are typically computed through some combination of numerical relativity (numerical solution of Einstein's equations) and semi-analytical models (e.g., post-Newtonian expansions matched to numerical relativity waveforms).  Noise properties are generally inferred from nearby, off-source detector data.  While a variety of tools are used to explore the multi-dimensional and possibly multi-modal parameter space, these tools typically compete on computational efficiency but should converge to the same posteriors provided the same assumptions are used and the user waits for a sufficiently long time.  On the other hand, noise models can differ, particularly when it comes to analysing/removing the impact of noise artefacts that may overlap with events.  For example, see the discussion regarding GW200129, which may or may not be the only event publicly released so far with unambiguous evidence of misalignment between the component spins and the orbital angular momentum, leading to precession\cite{Hannam:2021,Payne:2022}.  Signal models are not perfectly accurate, leading to potential systematic biases in the inferred parameters.  Despite much hard work and very significant progress, it remains challenging to compute waveforms that simultaneously cover the full parameter space of mass ratios and spins, include higher harmonics, incorporate both precession and possible eccentricity, and remain highly accurate.  Analysts are therefore forced to choose which approximations to make, but this leaves some questions unanswered: should we conclude that GW150921 is a precessing binary from an analysis with waveforms that allow for precession but not eccentricity\cite{Miller:2024}, or an eccentric binary from an analysis with waveforms that allow for eccentricity but not precession\cite{RomeroShaw:2020GW190521}?  Prior choices on the parameters, which describe a priori assumptions and should therefore reflect an individual scientist's prior beliefs, will obviously impact the posteriors of individual events.  One may hope that the impact of these priors is reduced when many events are combined into population inference, but this is not necessarily the case, as discussed below. 

While the two questions above are either largely resolved already or at least appear to be along a clear pathway toward resolution, the third question is one that the community is still grappling with: how to infer the properties of an entire population from individual events?  In principle, there are well-developed frameworks for doing so in both data-driven and model-specific ways.  Here, data-driven generally implies minimal assumptions about the model.  You may want to think of something like a kernel density estimator (KDE), which convolves individual measurements with a kernel to get a smooth population distribution.  Of course, you are probably thinking of a classical KDE in one dimension, while gravitational-wave posteriors are multi-dimensional (two masses, three spin vector components for each mass, perhaps two eccentricity parameters, sky location, distance, inclination, etc.).  Individual measurements are uncertain.  Detectors impose selection effects, which must be accounted for when inferring the underlying population to avoid Malmquist bias.  Even a KDE requires the user to make assumptions about regularisation: how wide should the kernel be?  While all of these problems have been addressed in the literature to some extent, it is not clear that we can ever converge on a unique and well-justified solution.  Especially so given inevitable questions about the utility of a purely model-driven approach: even if one could create an unambiguous multi-dimensional representation of the true population of merging black holes, what would one do with this?  How does an astrophysicist interpret something which exists in the form of, say, coefficients of a neural network that has perfectly learned the population, but cannot explain to a human what the relevant features are and what they mean?

Inference from models provides an alternative direction.  However, all physically motivated forward models are surely flawed and incomplete at present.  While it may be tempting to directly infer, say, the parameters of binary population synthesis models, such models are unlikely to faithfully represent even the formation of black holes from isolated binaries, and they certainly would not include many other possible formation channels, such as dynamical formation in dense stellar environments, hierarchical triples, or gas-assisted formation in active galactic nuclei.  And even including models for all of these channels, on which there is still significant disagreement, leaves a host of other sources of uncertainty, such as the star formation history of the Universe and, in particular, the distribution of metallicity across this star-formation history\cite{Chruslinska:2019}.  The risks of model mis-specification, possibly resulting in meaningless inference, are rife here.

Perhaps phenomenological models provide the best path forward.  Simple models, like (broken) power laws, proved extremely useful in modelling other astrophysical populations, such as the initial mass function of stars.  These models are easy to interpret.   They are convenient: in the example of an initial mass function, it is trivial to compute, say, the fraction of stars leading to supernovae and the formation of compact objects by analytically integrating a power law.  Even if the models are not completely accurate, there is great utility in such approximations.  However, gravitational-wave measurements are presenting a challenge.  Very simple models of the mass distribution, such as single power laws, have already been shown to fail, requiring increasingly complicated constructions of power laws, Gaussian peaks, dips and breaks\cite{GWTC3:pop}, bringing to mind Ptolemy's attempts to model a geocentric universe with many layers of epicycles and deferents.  Worse, it is not even clear which ``mass'' distribution should be modelled: at low masses, when many cycles of the inspiral fall into the sensitive frequency band of the detectors, the chirp mass $M_c \equiv m_1^{3/5} m_2^{3/5} (m_1+m_2)^{-1/5}$ is very accurately measured, but at higher masses, where the merger and post-merger ringdown dominate, the total mass $m_1+m_2$ is measured better, while the individual masses $m_1$ and $m_2$ are rarely measured with high accuracy.   And perhaps most worryingly of all, modelling the distribution of masses requires one to make assumptions about the distributions of other correlated parameters such as spins, meaning that the impact of prior choices may not disappear even with growing data sets.  Even with phenomenological models, model mis-specification remains a very serious risk for population-level models.

So what can we say, given these concerns about having to rely on uncertain models to convert well-measured parameters such as the chirp mass into distributions of poorly measured parameters such as individual masses, as well as the need to account for the prevalence of things we rarely saw and know very little about (say, low-mass black holes) when inferring the full underlying population?  

We do, in fact, know quite a lot.  Gravitational-wave detections include a host of very exciting individual events.  For example, GW190814 signalled the merger of two compact objects with an extreme mass ratio: a black hole of a bit over 20 solar masses (M$_\odot$) and a compact objects of about 2.6 M$_\odot$\cite{GW190814}, which could be a neutron star but I think is far more likely to be a black hole.  Another event with a low-mass black hole component is GW230529: a neutron-star -- black hole merger, where the latter likely falls into a mass range where we have few confident electromagnetic observations, between 2.5 and 4.5 M$_\odot$\cite{GW230529}.  On the other hand, GW190521, which already came up in this discussion, likely included two very massive black holes of $\sim 70$ M$_\odot$\cite{GW190521}, perhaps in a mass range where we would not expect to find black holes formed from the collapse of single stars in the course of normal stellar evolution.   

Furthermore, multiple population analyses, both data-driven and model-reliant, agree that a significant number of detected merging black holes have chirp masses slightly below 10 M$_\odot$, with a likely drop-off at somewhat higher masses before another peak approaching 30 M$_\odot$, and that there is likely a decrease but not a complete absence of more massive black holes extending beyond 40 M$_\odot$.  However, care should be taken when formulating more precise versions of these intentionally vague statements.  For example, whether the apparent paucity in mergers with a chirp mass a little above 10 M$_\odot$ can be translated into a gap in individual masses depends on the assumptions one makes about the pairing function between the black hole components, i.e., the mass ratio distribution\cite{Adamcewicz:2024,GalaudageLamberts:2025}.

\section{What masses should merging black holes have?}

Let us now attempt to re-formulate the question from the viewpoint of a theoretical astrophysicist rather than a gravitational-wave data analyst.  What black hole masses should we expect, given our best theoretical models?   Alas, the answer to this question is also far from clear.

A crude estimate is that stars born with masses below roughly 8 M$_\odot$ end their lives as white dwarfs, those with initial masses above roughly 20 M$_\odot$ end up as black holes, and those in the range between $\approx 8$ and $\approx 20$ M$_\odot$ form neutron stars.  However, the reality is far more complicated (for a recent review, see \cite{Heger:2023}).  There may well be alternating regions of explodability, where supernovae happen, and complete collapse into black holes.  Even when a supernova explosion does happen, fallback accretion of part of the material may form a black hole with a reduced mass.  The metallicity --- mass fraction of elements heavier than helium in the gas from which the star is initially formed --- impacts both the amount of mass lost in winds (and thus the remnant mass) and stellar compactness (and thus explodability).   The mass transfer history of stars in binaries will obviously change the masses, but perhaps also the explodability.  Pair-instability supernovae and their pulsational cousins are expected to lead some very massive stars to explode without a remnant and others to lose a lot of mass in the last stages of their evolution, creating a gap in the mass distribution, but there is still debate about the exact mass range of this gap because of uncertain nuclear reaction rates.  The strict page limitations on this submission make it impossible to discuss and properly credit these challenges; my brief summary is contained in the first chapter of another recent review\cite{Mandel:2024HMXB}.

In any case, the masses of black holes in general are likely very different from the masses of black holes in merging black-hole binaries.  It is certainly true that black holes observed in different settings --- as dark lenses in microlensing observations\cite{WyrzykowskiMandel:2019}, as accretors in mass-transferring low-mass X-ray binaries or wind-fed high-mass X-ray binaries\cite{Farr:2010}, in detached binaries observed astrometrically\cite{ElBadry:2023a,Panuzzo:2024} and in merging black holes detected through gravitational waves --- have very distinct mass distributions.  These distinctions are partly due to observational selection effects\cite{Jonker:2021,FishbachKalogera:2021}, but \textit{evolutionary} selection effects, which determine the evolutionary history of black holes in different settings and thus their masses, must surely play a role.  

So how exactly do merging binary black holes form?  

On the one hand, we may expect them to form quite commonly from massive stars in massive binaries, especially since massive stars are found predominantly in binaries (and often in multiple-component systems) and often prefer close massive companions\cite{MoeDiStefano:2017}.  A simple estimate suggests that if a Kroupa\cite{Kroupa:2001} initial mass function is assumed for the primary (more massive) star in a binary while the mass ratio between the secondary and primary is assumed to be flat, and that if the threshold for the formation of a black hole is an initial mass above 20 M$_\odot$, then one binary black hole should be formed for every $\sim 1200$ M$_\odot$ of star formation in binaries.  For round count, let us assume one binary black hole for every 1500 M$_\odot$ of total star formation.  The local universe has about $1.5 \times 10^7$ M$_\odot$ of star formation per Gpc$^3$ per year\cite{MadauDickinson:2014}, so we should expect $\sim 10,000$ binary black holes to be formed per Gpc$^3$ per year at redshift $z=0$, and an order of magnitude more at the peak of star formation around $z \approx 2$.  This far exceeds the merger rate of $\sim 10$--100 binary black holes per Gpc$^3$ per year from gravitational-wave observations\cite{GWTC3:pop}.  

Of course, we might expect that some of these binaries produce black holes that are too far apart to merge within the age of the Universe.  Indeed, in order to merge through the emission of gravitational waves, a circular binary of two 30 M$_\odot$ black holes needs to be separated by no more than $\approx 50$ solar radii (R$_\odot$).  But if we assume that binaries are born with a log-uniform distribution of initial separations between 20 R$_\odot$ (any closer and 30 M$_\odot$ stars would be in mass transfer at birth) and $1000$ AU $\approx 2 \times 10^5$ R$_\odot$, we find that $\approx 10\%$ should be close enough to merge.  This still yields a local merger rate of $\sim 1000$ per Gpc$^3$ per year that is far in excess of observations.   What prevents most of these massive binaries from merging as black holes?  

Stars expand as they evolve.  Red supergiants may reach radii of many hundreds, perhaps thousands, of solar radii.   Asking the progenitors of two black holes to fit into a binary with a separation of a few tens of solar radii is worse than fitting a square peg into a round role -- it is trying to shove a peg into a hole that is orders of magnitude smaller than the peg.  The stars seem destined to merge as they age and expand.  Thus, perhaps the right question is not why the binary black hole merger rate is lower than our naive estimate, it is why binary black holes merge at all.

Most of the source modelling in gravitational-wave astronomy centres on various strategies to overcome this challenge of bringing two black holes close enough to allow them to merge within 14 Gyr or less, but without having the stars merge prematurely, before two black holes are formed (for reviews with many more citations, see\cite{MandelFarmer:2018,Mapelli:2021,MandelBroekgaarden:2021}).  A binary that is initially sufficiently wide to accommodate expanding stars but not quite wide enough to fit them at their full extent will experience mass transfer; however, that mass transfer does not always lead to merger, and may sometimes bring the binary closer together just when the stars are stripped of their envelopes and reduce in radii so that they can fit into a tighter system.  Dynamically unstable mass transfer, known as common-envelope evolution, may or may not play a particularly important role for forming binary black holes.  Chemically homogenous evolution of rapidly rotating, fully mixed stars in initially very tight binaries could suppress stellar expansion and allow black hole formation in situ.  Black holes could be brought into binaries and the binaries tightened by dynamical gravitational interactions, either in dense environments such as stellar clusters or in hierarchical triples in which the inner binary is driven to a higher eccentricity by the tertiary companion.  Stellar-mass black holes embedded in the accretion disks around supermassive black holes in active galactic nuclei may be caught in migration traps, where mergers are potentially assisted by gas drag and/or accretion.   

Depending on the channel at play, a variety of tools have been deployed to evaluate the contribution of the channel to the binary black hole merger rate and predict the black hole mass distribution.  These include detailed three-dimensional hydrodynamical models of individual key phases of stellar or binary evolution, such as supernovae and common envelopes; one-dimensional models of stellar evolution and stellar winds; binary population synthesis based on recipes from such detailed simulations to rapidly Monte Carlo sample a large population of binaries; dynamical models of star clusters, nuclear clusters, and globular clusters of varying degrees of complexity; joint gas and dynamical modelling of active galactic nuclei; and a broad range of semi-analytical or phenomenological models and approximations of varying degrees of sophistication and fidelity.

However, our understanding of many of the key aspects of these formation channels that impact the merging black hole masses and rates (and thus the relative importance of various channels) is still insufficient to make confident statements, despite very significant efforts and some notable advances over the past decades.  A brief and incomplete list of the key uncertainties includes wind mass loss, mass transfer, and collapse physics including remnant masses and natal kicks for all formation channels; initial cluster properties for dynamical formation; the very viability of chemically homogeneous evolution and the range of rotation rates and metallicities at which it may operate for the relevant channel; initial conditions of triples for the hierarchical triple channel; detailed physics of migration traps for the active galactic nuclei channel; and mass transfer physics for the isolated binary evolution channel which relies on mass transfer to shrink the binary.  The significant impacts of different assumptions for the latter channel on the predicted mass distribution are illustrated, for example, in figure 4 of reference\cite{Broekgaarden:2022}.

\section{How do we make progress?}

Given the remaining challenges in directly interpreting the mass distribution of merging binary black holes, how can we move forward? 

Merging binary black holes represent the very last stages in the lives of stars and binaries.  The reason interpreting them is so challenging is that so much happens during their lives between birth and death that we are still unsure about: the full history of stellar evolution including mass loss and core collapse, possible mass transfer, and possible tidal or dynamical interactions.  

Furthermore, because of the poor spatial resolution of gravitational-wave detectors, we do not know about the environments of these compact objects --- whether they are in star clusters or in galactic fields, what the likely progenitor metallicity is, how long the merger is delayed relative to star formation in the host galaxy.  At least not unless we are exceptionally fortunate to detect an electromagnetic counterpart, which so far happened for one gravitational-wave neutron star merger\cite{GW170817:MMA} and never confidently for binary black hole mergers, despite several intriguing candidates.   

Instead of relying on the final death mask taken at merger to reconstruct the entire preceding lifetime of black holes, we can learn far more by observing black holes, and the evolution of massive stellar binaries more generally, in a range of settings and at various evolutionary stages and comparing these snapshots to extract the full evolutionary history.  Some examples have been mentioned above, and I highlight a few particularly promising ones here.

Microlensing by black holes combined with precise astrometry has the potential to simultaneously yield black-hole masses and velocities of apparently single black holes, although this technique has only been demonstrated for one source so far\cite{Sahu:2022}.  

Black-hole high-mass X-ray binaries such as Cygnus X-1 can serve as exquisite probes of mass transfer physics, including the stability and conservativeness of mass transfer, the angular momentum lost during non-conservative mass transfer, and wind Roche lobe overflow.  The current state of play in this area is discussed in more detail in\cite{Mandel:2024HMXB}.  Meanwhile, black-hole low-mass X-ray binaries can help us understand what happens during a common-envelope phase which likely occurred during mass transfer from the black hole's progenitor onto its low-mass companion\cite{HiraiMandel:2022}, as well as shed light on stellar rotation and rotationally-limited accretion.  Both types of X-ray binaries are potential probes of black-hole spins and spin-orbit misalignments\cite{Poutanen:2021}, although existing measurements of black-hole spins with electromagnetic observations may suffer from serious systematic biases\cite{MillerMiller:2015}.  

Detached binaries containing black holes make it possible to measure the mass lost during core collapse and natal kicks\cite{VignaGomez:2024}.  

While we are limited to the environment of the Milky Way and other nearby galaxies for these kinds of observations, we can often directly measure the metallicity of the stellar companion to the black hole and explore the local star formation history and the likely importance of any dynamical interactions.

These snapshots must be supplemented with a broad range of observations of non-black-hole systems, from luminous red novae associated with common envelope events and post-mass transfer binaries, to winds from massive stars and supernova explosions.  We expect the available data sets for many of these observations to explode in the coming years thanks to new instruments and surveys, such as the Vera Rubin Observatory Legacy Survey of Space and Time (LSST)\cite{Howitt:2020}.  Taken together, and interpreted with the aid of the host of modelling tools at our disposal, these may finally allow us to create a concordance model of binary evolution and take full advantage of gravitational-wave measurements, but it is a massive undertaking indeed.

\section*{Acknowledgments}

I acknowledge support by the Australian Research Council Centre of Excellence for Gravitational Wave Discovery (OzGrav) through project number CE230100016.  I am grateful to Reinhold Willcox for comments on the manuscript.

\section*{References}
\bibliography{Mandel}

\end{document}